\documentclass[printer]{aa} %

\usepackage{graphicx}
\usepackage{txfonts}
\usepackage{natbib}
\usepackage{hyperref}

\newcommand{\spwat}{H_{2}O}
\newcommand{\snit}{N_{2}O}
\newcommand{\snita}{NO_{2}}
\newcommand{\spox}{O_{2}}
\newcommand{\spoz}{O_{3}}
\newcommand{\smeth}{CH_{4}}
\newcommand{\scarb}{CO_{2}}

\begin{document}

\title{Transmission of the AtmosPhere for AStronomical data: TAPAS upgrade}

\author{R. Lallement\inst{1}
\and J.~L. Bertaux\inst{2}
\and S. Ferron\inst{3}
\and C. Boonne\inst{4}
\and E. Richard\inst{4}
\and F. Lefèvre\inst{2}
\and J.V. Smoker\inst{5}}

\institute{LIRA, Observatoire de Paris, PSL University, CNRS, 5 Place Jules Janssen, 92190 Meudon, France\\
       \email{rosine.lallement@obspm.fr}
\and 
LATMOS, Sorbonne-Université, Paris, France
\and 
ACRI-ST, 260 Route du Pin Montard, BP234, 06904 Sophia-Antipolis, France 
\and 
Institut Pierre Simon Laplace (IPSL), AERIS data centre, 75252 Paris, France 
\and 
European Southern Observatory, Alonso de Cordova 3107, Vitacura, Santiago, Chile}

\date{Received ; accepted }
\titlerunning{Tapas}
 
\abstract
{State-of-the-art molecular databases and realistic global atmospheric models allow to predict accurate atmospheric transmittance spectra. Observers with ground-based spectrographs may use this information to identify the telluric absorption lines, to correct, fully or partially,  their astronomical spectra for those lines, or take them into account in forward models.}
{The TAPAS online service provides atmospheric transmittance spectra of the most important species as well as Rayleigh extinction, adapted to any observing location, date, and direction. We describe recent updates, improvements, and additional tools.}
{TAPAS is interpolating in location in the atmospheric profiles of temperature, pressure, $\spwat$, $\spox$ and $\spoz$ that are extracted from the meteorological field of the European Center for Medium Term Weather Forecast (ECMWF) for the date and time of the observation. The composite profiles are produced by a Data Terra/AERIS/ESPRI product called Arletty, and supplemented by auxiliary climatological models for additional species.  The transmittance spectra are computed with the LBLRTM code. The default width of the spectral pixels is chosen to ensure that the shapes of all the absorption lines are reproduced for each species. Major improvements with respect to the previous TAPAS are: -  the extension of the wavelength range in the near-UV down to 300 nm; the extension in the near-IR up to 3500 nm; - the use of the recent version of the HITRAN database (HITRAN2020); - the addition of $\snita$ transmittance, to complement $\spwat$, $\spox$, $\spoz$, $\snit$, $\scarb$ and $\smeth$ ;  - an increased accessibility and a reduced time to obtain the results; -the possibility to force the total $\spwat$ column to match the one measured at the observatory at the time of record.}
{We show $\spoz$ absorption in the near-UV and near-IR and $\snita$ absorption in the visible. We illustrate the quality of TAPAS by means of comparisons between models and ESO/VLT/CRIRES recorded spectra of a hot star with spectral resolution $\sim$130,000, in two intervals in the near-IR with strong $\spwat$, $\snit$, $\scarb$ and $\smeth$ absorption. We describe the measurement of an instrumental Line Spread Function (LSF) based on TAPAS $\spox$ lines and a method using the Singular Value Decomposition technique that can be made entirely automated.}
{The new TAPAS tool provides realistic simulations of the telluric lines. It gives access to the weakest $\spwat$ or $\spox$ lines, and to the very weak, highly irregular $\snita$ lines. It can be used to improve the wavelength assignment when calibration lamps provide few emission lines, and to measure accurately the LSF in most regions where telluric features are present.  The extended wavelength range will be particularly useful for future or recent spectrographs in the near-UV, and in the near-infrared.}


\maketitle

\section{TAPAS description and evolution from the previous version}\label{intro}

Telluric absorption is a contaminant of astronomical spectra that must be taken into account and, ideally, eliminated from the data. As a result of the tremendous improvements made over the last few decades, in the meteorological fields on the one hand and of molecular data on the other, it has become possible to perform this correction based on a predicted model of the telluric transmittance spectrum, replacing the division by the spectrum of a smooth continuum comparison star, recorded under the same conditions as those of the target of interest \citep[see][for a first application with the GEISA spectroscopic database]{Lallement93}. This procedure saves observing time and avoids features present in the spectrum of the comparison star. To this end, the free service TAPAS was launched in 2012 and online in 2014 \citep[see ][for details]{Bertaux2014}. During the last decades, telluric corrections became more frequent, sometimes routinely done, and new wavelength ranges contaminated by telluric features could be used. Observations which require especially accurate telluric corrections are those dedicated to exo-planets. In the case of radial velocity measurements, residuals of insufficiently corrected telluric lines, or undetected so-called {\it micro-tellurics} perturb the periodic signals due to planets. In the case of transit spectroscopy, which aims at detecting an atmosphere around an exo-planet, and identifying the presence of gases, like $\scarb$ and $\spwat$, the correction is of crucial importance. As a matter of fact, the tiny circular atmosphere around the exo-planet is giving a spectral signature, as a tiny change of the host star spectrum, and very precise measurements of the much stronger telluric features variations are mandatory.  

\subsection{Short description}
TAPAS\footnote{https://tapas.aeris-data.fr/en/home/} simulates with exquisite detail the atmospheric transmission spectra due to the main absorbing species as a function of date, observing site, and either zenith angle or target celestial coordinates.  TAPAS uses the Arletty facility from the "Ensemble de Services pour la Recherche à l’IPSL" (ESPRI) Data and Services Center from the "Institut Pierre-Simon Laplace" (IPSL), part of the Data Terra/AERIS  (a French Atmospheric Chemistry Data Center) infrastructure. Arletty \citep{Hauchecorne99} computes vertical pressure, temperature, and $\spwat$, $\spox$ and $\spoz$ profiles extracted from the extremely detailed model of the global atmosphere produced by the "European Centre for Medium-Range Weather Forecasts" (ECMWF) using a grid of 0.5 degree in longitude and latitude. The ECMWF profiles are refreshed every 6 hours through assimilation of new data from space and ground. Arletty selects the model closest in time to the observer's request and interpolates in location within the ECMWF grid. Additional models based on data are used to calculate the distributions of $\snit$, $\scarb$ and $\smeth$. For each date, TAPAS interpolates in the measurements of the annual evolution of the mean abundance of those species to select the appropriate value of the relative abundance to dry air (or mixing ratio), and then it distributes in altitude the concentration of the species following this fixed mixing ratio and the pressure and temperature altitude profiles.

Atmospheric transmittance spectra are integrated from the altitude of 70 km down to the observatory in each of the altitude layers of ECMWF, and above 70 km in the climate model MSIS-90. They are based on molecular data from the HITRAN database \citep{Rothman2009, Gordon2017, GORDON2022} and computed by means of the Line-By-Line Radiative Transfer Model (LBLRTM ) \citep{Clough1995, Clough2005}. For each species and each altitude layer, the local temperature and pressure are used to compute air- and self-broadening, as well as pressure line shift for each HITRAN transition. TAPAS calculates the transmittance separately for each isotope present in HITRAN, using standard terrestrial abundance ratios. For the basic mode of the TAPAS output, the grid spacing is automatically chosen to be small enough to allow each atmospheric line from each species to be accurately represented. The resulting grid steps are on the order of 1 m\AA~ (Resolving power R $\simeq$ 5x10$^{6}$ at 1000 nm, assuming a resolving power element is sampled by two grid steps). This is important because this opens the possibility to reproduce their exact shapes and determine the instrumental function of the spectrographs (see Section \ref{findpsf}). The transmission spectra may be obtained separately for individual gases. The spectra can be transformed in the observer's frame or in the solar system barycentric frame, and can be convolved to adapt to any spectral resolution.

\begin{figure*}
\centering 
\includegraphics[width=0.98\textwidth,height=9cm]{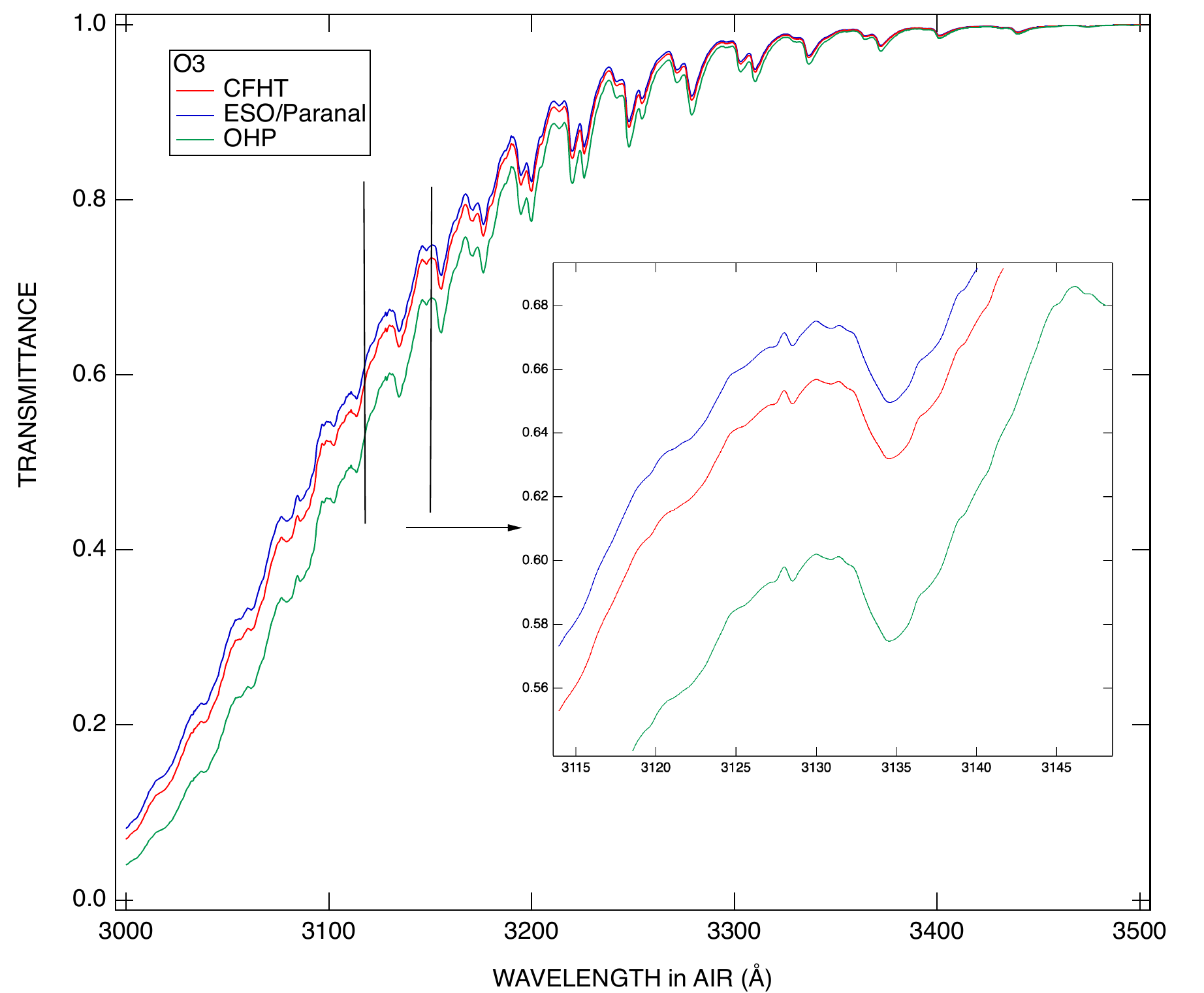}
\caption{Example of transmittance spectrum of $\spoz$ in the near-UV. A zoom on one of the intervals is inserted, to show some of the very small details.}
\label{Fig:O3_NUV}
\end{figure*}

\begin{figure*}
\centering 
\includegraphics[width=0.98\textwidth, height=10cm]{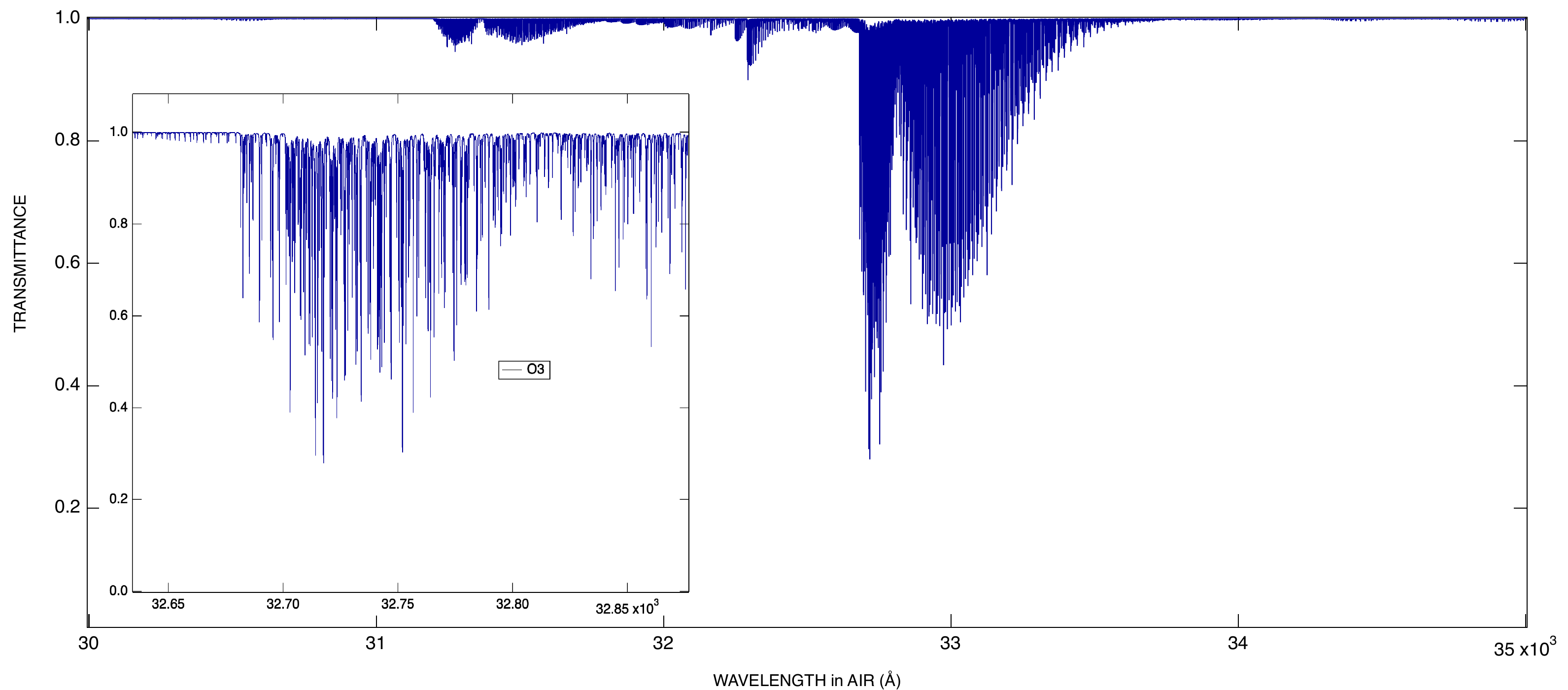}
\caption{Example of transmittance spectrum of $\spoz$ in the near-infrared domain. A zoom in one of the strongly absorbed spectral regions is inserted.}
\label{Fig:O3_NIR}
\end{figure*}

\subsection{TAPAS limitations and advantages, telluric correction techniques}

TAPAS does not perform itself a removal of telluric features in user's recorded spectra, at variance with, e.g. the ESO software MolecFit \cite{Smette2015} or, e.g., the Telfit tool \citep{Gullikson2014}. Instead, it provides the necessary detailed information to perform a correction, without the need for a local software installation. As far as transmittance spectra are concerned, the main difference between TAPAS and the above mentioned tools is the choice of the global atmospheric model, the European ECMWF for TAPAS, and the US "Global Data Assimilation System" GDAS model for the two other tools.  TAPAS serves as a basis for line identifications or for various post-processing correction techniques.  It allows one to obtain easily, rapidly, and online the best transmittance models, species by species, to be subsequently used by the user in various ways. It allows one to identify which feature is of telluric origin, to predict the telluric contamination of planned observations, and to use the transmittance spectra in locally developed codes of telluric removal or in forward models combining telluric and astronomical information. 

Several simple or sophisticated techniques have been already implemented by different groups to use modeled atmospheric transmittance spectra to correct astronomical spectra for telluric contamination. The most simple techniques adjust the model to the data by varying the instrumental function of the spectrograph, the wavelength offset between the data and the model, if necessary, and the columns of absorbers A$_{i}$ (i=1,N for N different absorbers). This latter adjustment makes use of N free parameters X$_{i}$ and T$^{X_{i}}(A_{i}$), which is the model transmission T($A_{i}$) elevated to the power X$_{i}$, corresponding to an increase of the column of the absorber A$_{i}$ by a factor X$_{i}$. In practice, TAPAS predicts very accurate transmissions for all species except $\spwat$, and in this case N=1 (see Section \ref{spectra}). Because the integrated water vapor varies on timescales of hours or even minutes, there are no accurate predictions of the water vapor transmittance. One may use local measurements based on a radiometer \citep{Smette20}, however, at the time of writing, the retrieval algorithms used by the radiometers at Paranal tend to underestimate the amount of precipitable water vapor for values below $\simeq$ 1mm \citep[see also][]{Ivanova23}. In the case of weak to moderate absorption, a very simple way to find a value of X$_{i}$ best fitting the data and to adjust the instrumental function and the wavelength offset is to use as a criterion the minimization of the length of the spectrum obtained after division of the data by the model. This "rope-length" method uses the fact that for the optimal model there are no irregular residuals whose effect is a significant increase of the length of the spectrum \citep[e.g.][]{Cox17, CamiMessenger18, Elyajouri18}. For stronger lines, another method is to adjust the data to a forward model based on the product of the telluric transmittance by another function, which can be a continuum \citep[e.g.][]{Ivanova23},  the product of a continuum and an absorption band \citep[e.g.][]{Puspi13}, the product of a continuum, a modeled stellar spectrum and an absorption band \citep[e.g.][]{Puspi15}, or other functions. Indeed, such a forward modeling technique is necessary in the case of cool stars whose spectra contain numerous deep and narrow lines. Intermediate methods using 
both "rope-length" minimization and forward modeling have also been proposed \citep{CamiMessenger18}. Finally, sophisticated adjustments by means of chi square method are used by Molecfit \cite{Smette2015}. They include the possibility to vary, in addition to wavelength offset, spectral resolution and abundances, the shape of the instrumental function. In the case of very strong lines, in particular in the red or near-infrared domain, the correction for telluric lines becomes extremely sensitive to very small discrepancies between the model and the data.  Modifications of the downloaded transmittance may be performed to remove the residuals due to such small discrepancies. This is the case, for example, of the APERO technique developed by \cite{Cook22} and used for CFHT/SPIROU observations. APERO uses the residuals extracted from the spectra of hot stars initially corrected based on TAPAS transmittance spectra to produce a database of corrections \citep[e.g.][]{Artigau14,Artigau21} and perform a refined PCA-based correction.  

\begin{figure*}
\centering 
\includegraphics[width=0.98\textwidth,height=10cm]{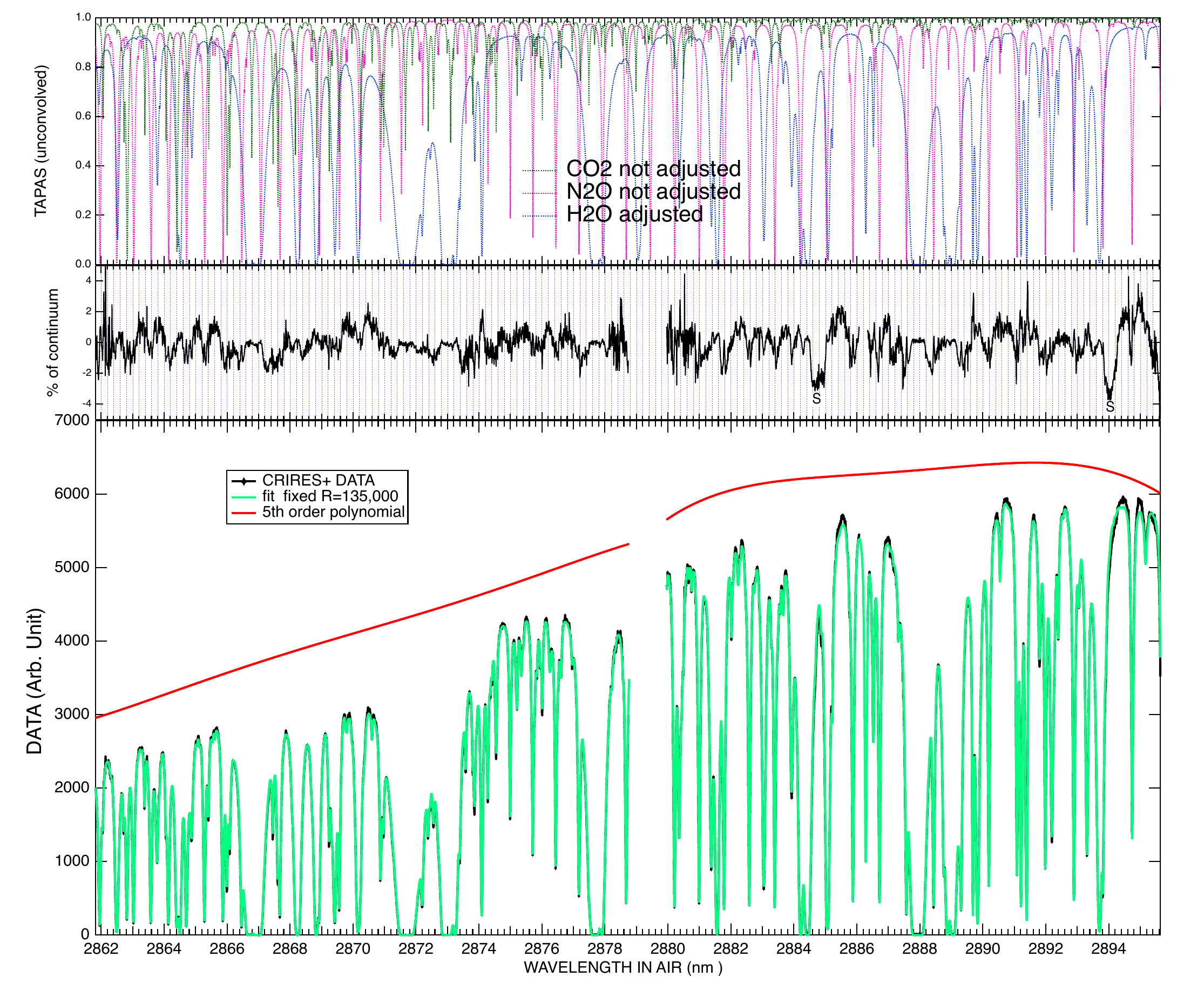}
\caption{Example of comparison between TAPAS predictions and a CRIRES spectrum of the hot star $\beta$ CMa. Top: The three transmittance spectra of $\spwat$, $\snit$ and $\scarb$ computed by TAPAS and before convolution are shown (right scale). Bottom: Data (in black) are fitted to the product of their convolved product by a 5th order polynomial (green curve). The fitted polynomial is displayed separately (red curve). The central gap in the data is due to the transition between the second and the third detector of the instrument. Middle: Differences between data and model expressed in percentage of the fitted continuum. The two broad and shallow features marked by S cannot be residual telluric features and are very likely stellar.}
\label{Fig: BetaCMa_1}
\end{figure*}

\begin{figure*}
\centering 
\includegraphics[width=0.98\textwidth,height=10cm]{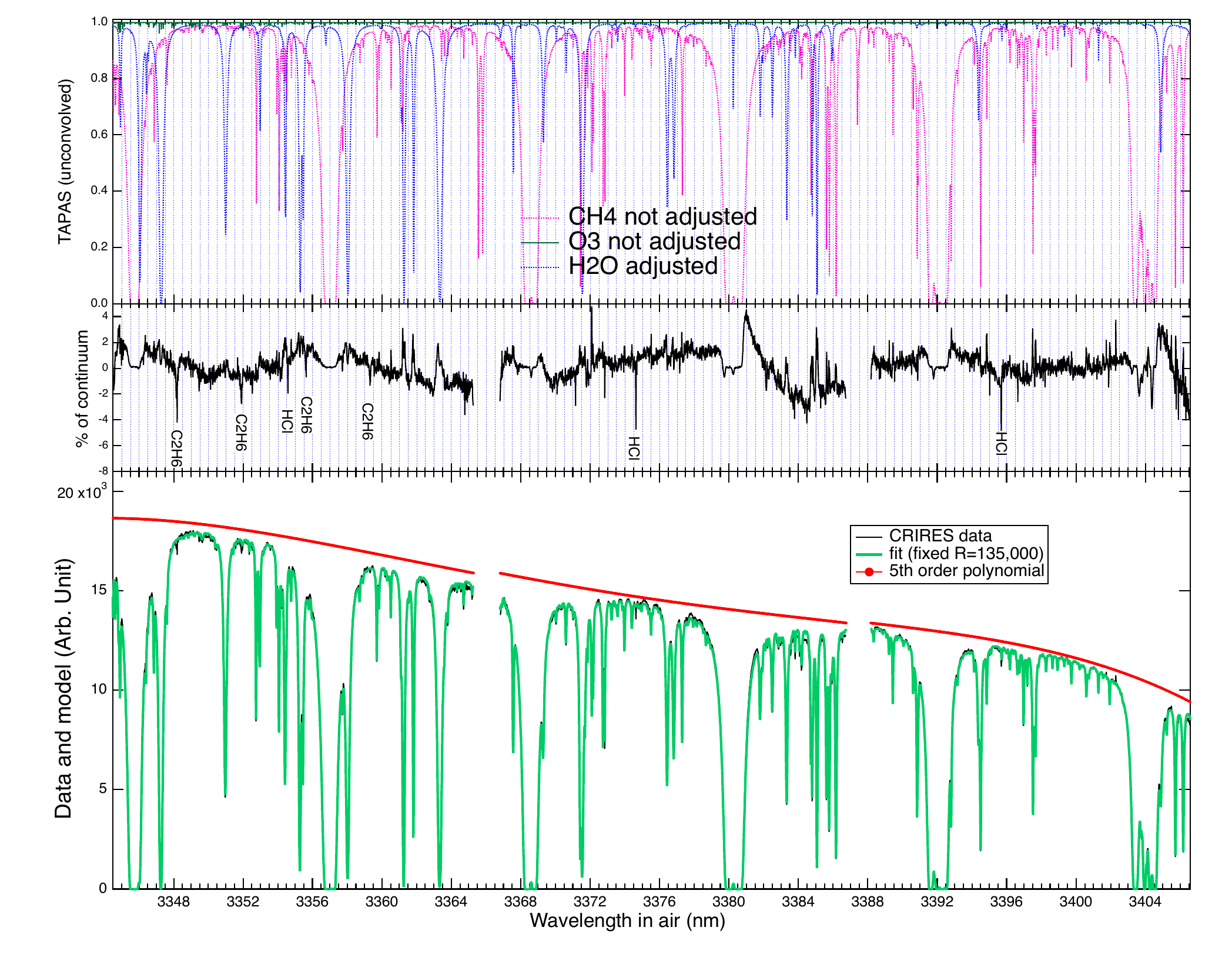}
\caption{Same as Fig. \ref{Fig: BetaCMa_1} for a different spectral interval characterized by absorption lines of  $\spwat$, $\smeth$ and $\spoz$. Top: TAPAS transmittance spectra are shown at top (right scale). Bottom: Data (in black) are fitted to the product of their convolved product by a 5th order polynomial (green curve). The fitted polynomial is displayed separately (red curve). The two data gaps are due to the transitions between the detectors of the instrument. Middle: Differences between data and model expressed in percentage of the fitted continuum. A few  weak and sharp absorption lines are absent from the model and were identified as due to C$_{2}$H$_{6}$ and HCl (see text).}
\label{Fig: BetaCMa_2}
\end{figure*}

\begin{figure}
\includegraphics[width=0.49 \textwidth]{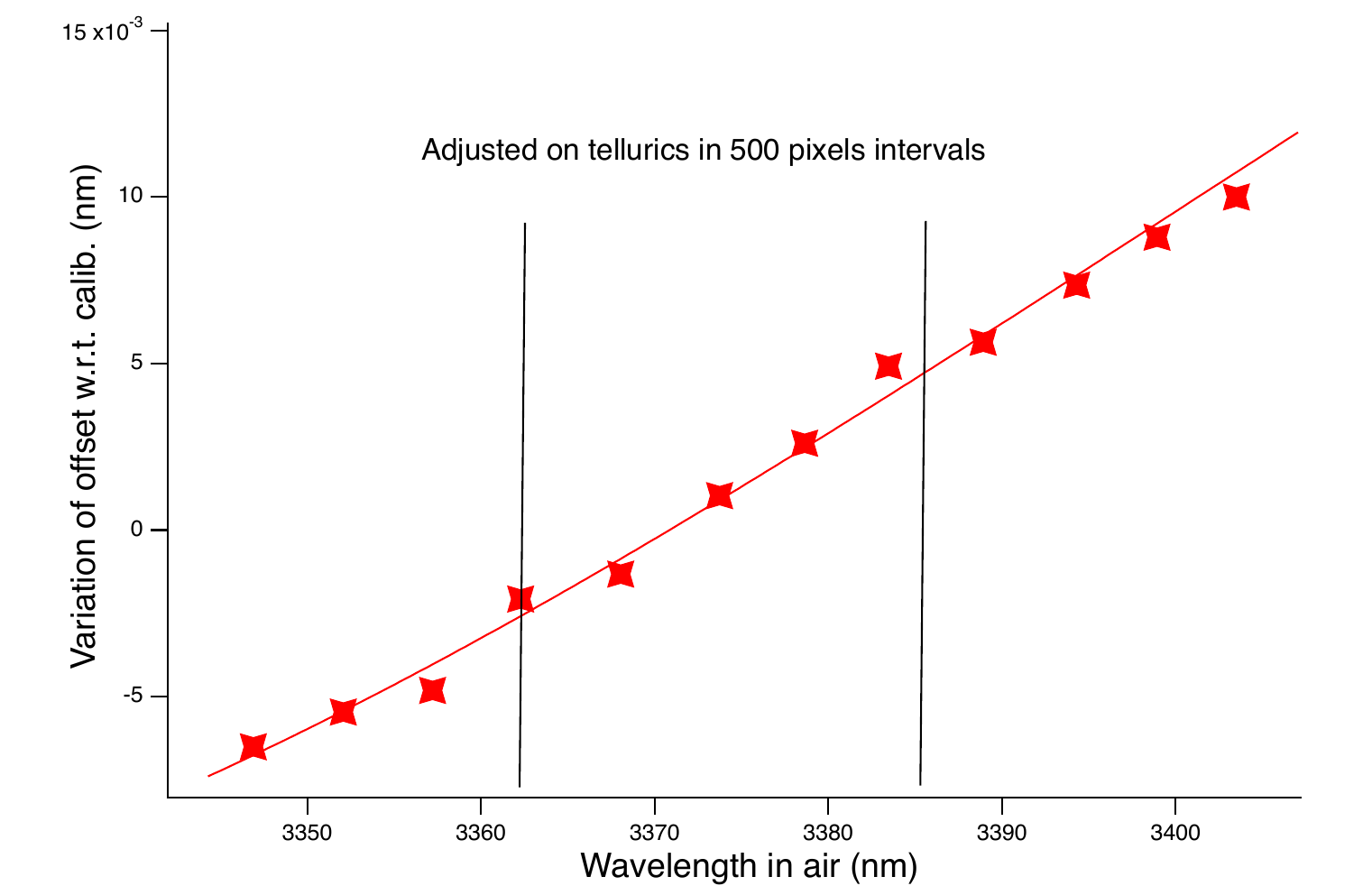}
\caption{Evolution of the difference between initially assigned wavelengths and refined telluric-based wavelengths, in the case of data from Fig. \ref{Fig: BetaCMa_2} and in 500 pixels intervals. The average wavelength offset has been removed. The black vertical lines show the locations of the gaps between the detectors. The red line is the polynomial used in the model shown in Fig. \ref{Fig: BetaCMa_2}.}
\label{Fig:shift_evol}
\end{figure}

\begin{figure*}
\centering 
\includegraphics[width=0.9\textwidth, height=7cm]{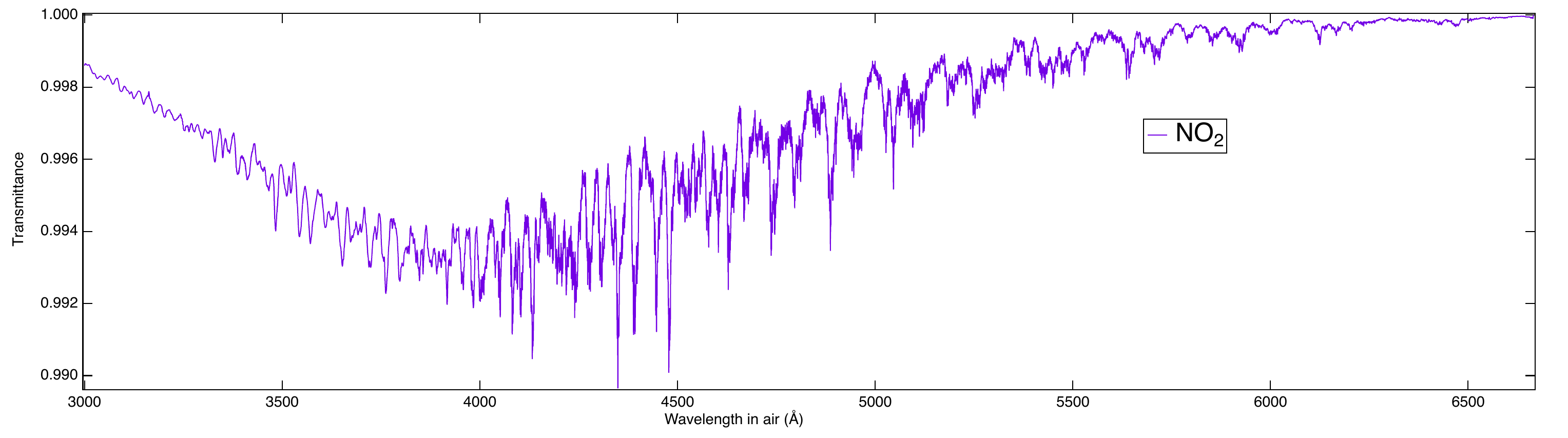}
\caption{Example of a transmittance spectrum of $\snita$, for ESO-Paranal, in May, and for an airmass of 2. Absorption depths are very small, however, the irregular pattern makes this absorbing species difficult to identify.}
\label{Fig:NO2}
\end{figure*}

\subsection{TAPAS upgrade}

The TAPAS facility has recently been updated, and the purpose of this document is to illustrate this evolution. 
The main type of improvement concerns the computations of the transmittance spectra. It includes an important extension of the wavelength range, in the near-UV (NUV) from 350 nm down to 300 nm, and, in the near-infrared (NIR), from 2500 nm up to 3500 nm. 
This may be useful in the context of the development of a new generation of spectrographs in these ranges, namely the  Cassegrain U-Band Efficient Spectrograph  (CUBES) at ESO/VLT \citep{Cristiani22} in the NUV, and CRIRES at the ESO/VLT \citep{Dorn23} in the NIR. 
Updates were made on the models used to calculate the distributions of $\snit$, $\scarb$ and $\smeth$. For dates before 2002, data were extracted from the "World Data Center for Greenhouse Gases" \footnote{https://https://gaw.kishou.go.jp/}
 and  the "Global Monitoring Laboratory" \footnote{https://gml.noaa.gov/}. For dates from 2002 and later, TAPAS uses the most recent values from the Copernicus Atmosphere Monitoring Service (CAMS) \footnote{https://atmosphere.copernicus.eu/}. Updates also include the addition of the $\snita$ transmittance, to complement $\spwat$, $\spox$, $\spoz$, $\snit$, $\scarb$ and $\smeth$. Moreover, there is now the possibility for the user to force the $\spwat$ model to the humidity conditions measured at the observatory at the time of record. This may be useful in the case of instruments reaching the NIR domain such as SPIRou at the Canada France Hawaii Telescope \citep{Donati20}, GIANO at the TNG telescope \citep{Origlia14}, CARMENES at the Calar Alto Observatory \citep{Quirrenbach12} and the NIRPS at the ESO/La Silla 3.6-metre telescope \citep{Wildi17}.  Last but not least, another improvement is the use of the most recent version of the HITRAN database (HITRAN2020).  
On the other hand, the TAPAS interface has also been substantially improved. It contains additional explanations to facilitate the user's choices, and additional information as part of the downloaded spectra. It allows several formats for the retrieved models, and the possibility to save parameters in case of series of similar requests. Finally, a last type of update concerns the atmospheric model. TAPAS benefits from the constant improvements of the ECMWF model. One of its main changes is the increase number of altitude layers from 70 to 137. The expected differences, here in the case of the telluric transmittance spectra for ground-based observations, are very small, because the altitude coverage was already very good at low altitudes where most of the absorption lines are built. Still, it can only improve the line shapes.

In Section \ref{spectra} we show transmittance spectra in the new wavelength intervals. In the case of the near-infrared domain, we compare the TAPAS predictions with high-quality, high-signal spectra of a hot star recorded with the CRyogenic InfraRed Echelle Spectrograph (CRIRES) at the ESO/VLT \citep{Dorn23}.  
In Section \ref{findpsf} we show how TAPAS transmittance spectra can be used to determine the instrumental Line Spread function (LSF) of a spectrograph. We used the technique based on the Singular-Value Decomposition (SVD) of the matrix linking the LSF to the data. In Section \ref{usingtapas}, we describe the online facility.

\begin{figure}
\centering 
\includegraphics[width=0.49\textwidth, height=10cm]{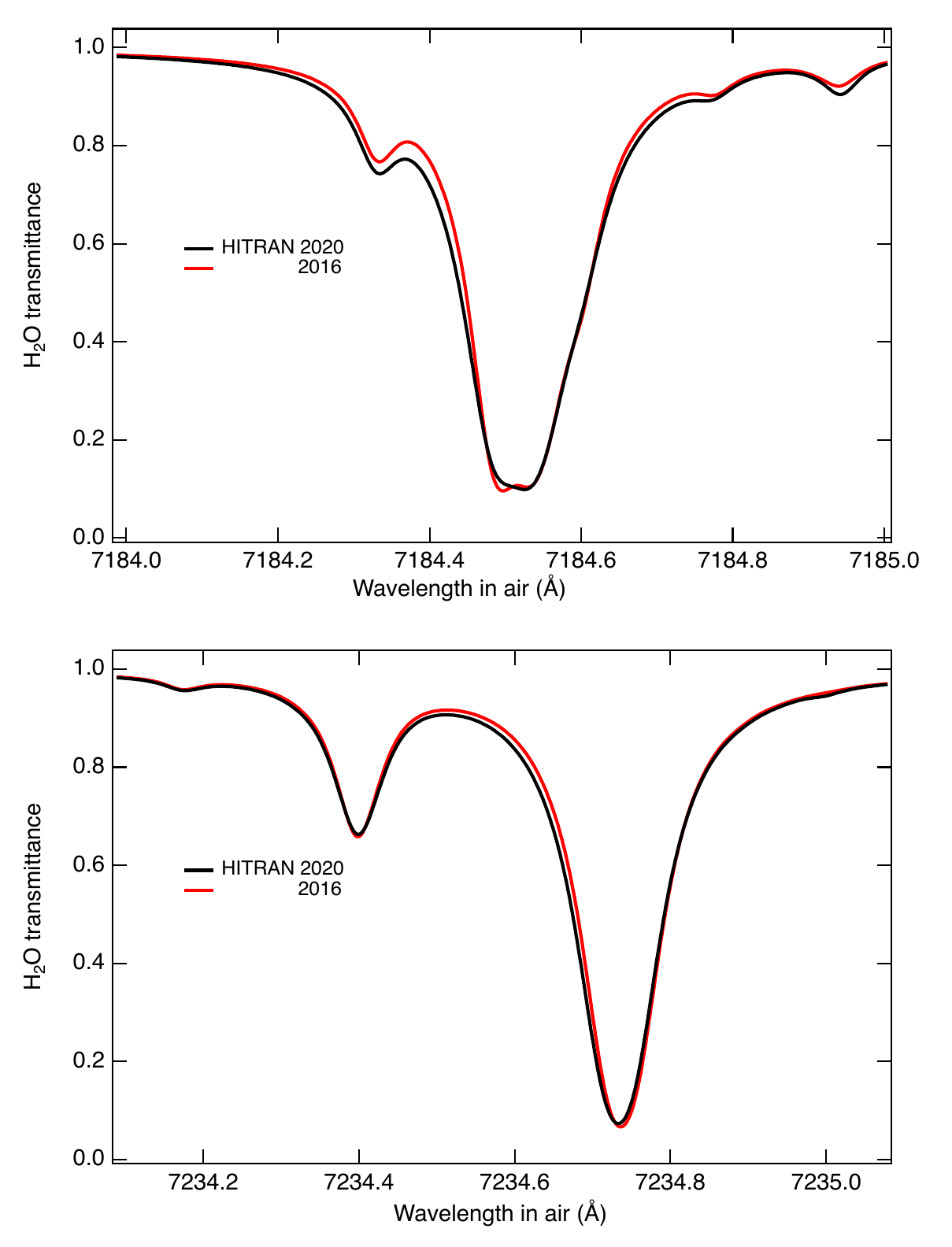}
\caption{Examples of changes from HITRAN 2016 to HITRAN2020. We use two of the small discrepancies between data and model detected  by \cite{Allart22} in the case of $\spwat$ lines, based on ESPRESSO spectra. The P-Cygni profiles found by the authors indicate that the deep part of the modeled line at 7184.6 \AA~ was slightly blue-shifted with respect to the data (see the double structure at bottom in the top plot which results in a global blue-shift) while it was the opposite in the spectral regions shown in the bottom figure. The new HITRAN prediction shifts the lines in a way to correct for these effects. Also, note the new weak line predicted by HITRAN 2020 at 7235 \AA~ that was absent in HITRAN 2016.}
\label{Fig:HITRAN16_20}
\end{figure}

\section{New transmittance spectra}\label{spectra}

\subsection{$\spoz$ absorption in the NUV and NIR}
The extension of the TAPAS domain opens new perspectives. The addition of the 300-350 nm interval allows us to obtain very precise transmittance spectra of $\spoz$, a strong absorber in this area. These models, which can be easily obtained for any period of time and conditions of observation may be useful, e.g., to simulate data from the future VLT/CUBES spectrograph of ESO. We show in Fig. \ref{Fig:O3_NUV} the absorption by $\spoz$, here, as an example, for ESO Paranal in May at zenith. The absorption pattern is very complex, and features of very different spectral widths are present, as illustrated in a zoomed interval. In the NIR, the addition of the 2500 to 3500 nm interval allows us to simulate data from spectrographs such as CRIRES. We show in Fig \ref{Fig:O3_NIR} the $\spoz$ absorption bands between 3000 and 3400 nm for ESO Paranal in May for an airmass of 2. 

\subsection{$\spwat$, $\snit$, $\scarb$, $\smeth$ in the NIR}
We illustrate the quality of the tool with two examples of comparisons between TAPAS predictions and recorded CRIRES spectra of the hot star $\beta$ CMa in two different spectral intervals in the L band (2862 to 2897 nm and 3346 to 3406 nm respectively). The setting used was L3377 with a slit width of 0.2". Figs. \ref{Fig: BetaCMa_1} and \ref{Fig: BetaCMa_2}  show predicted transmittance spectra appropriate to Paranal, the date of observation, the direction of the star, for $\spwat$, $\snit$ and $\scarb$ and $\spwat$, $\smeth$ and $\spoz$ respectively. The water vapor column measured at the observatory and indicated along with the data was imposed, and subsequently refined by elevating the transmittance at an adjusted power. Here for the exceptional condition of low humidity, and a radiometer predicted column of 0.51 mm at the zenith, we had to elevate the TAPAS output  $\spwat$ transmittance at a power 1.3 (i.e. increasing the column by 30$\%$). Other species were kept unchanged. The products of the transmittance spectra for the relevant absorbing species were convolved with a Gaussian instrumental function and data were fitted to the product of this resulting spectrum by a fifth order polynomial function representing the stellar spectrum as seen by the spectrograph, whose coefficients were let free to vary. Prior to this adjustment, the offset in wavelength between the model and the data was computed all along the spectral bands, based on the telluric features, assuming they provide the optimal calibration. The introduction of such an offset is made necessary by the lack of features in the spectra of the calibration lamps in these CRIRES spectral regions, and the resulting uncertain wavelength assignment. To do so, single-value offsets were firstly determined in consecutive intervals of 500 pixels, and the resulting series of measured offsets was subsequently fitted to a second order polynomial function of the pixel number, as illustrated in Fig. \ref{Fig:shift_evol} for the wavelength interval of Fig. \ref{Fig: BetaCMa_2}. This function, representing the wavelength difference between the value issued from the calibration and the one issued from the telluric features was subsequently imposed in the data-model final adjustment.

Because only telluric features are shaping the spectrum of this target star, characterized by a smooth continuum, it was possible to appreciate the excellent quality of the predictions (see Figs. \ref{Fig: BetaCMa_1} and \ref{Fig: BetaCMa_2}). A careful inspection of the correspondence between data and the TAPAS model revealed only a few very weak and narrow features in the second interval that are absent in the model. Those lines are due to C$_{2}$H$_{6}$. C$_{2}$H$_{6}$ and HCN produce a few very weak absorptions in the 3000-3500 nm domain and will be the subject of future work. During the continuum adjustment process, we initially imposed a fixed resolving power R of 90,000, but this value appeared as significantly too low. In order to obtain a good fit, we had to increase R up to 130,000. This is linked to the use of adaptive optics (see below). 

There are some important aspects of the use of models such as TAPAS in the case of instruments like CRIRES. In this spectral domain, very few emission lines from calibration lamps can be used, resulting in a poor wavelength calibration.   The number of telluric features being very important in this spectral domain, and telluric lines being distributed over the whole spectral intervals we considered, it was possible to use the TAPAS transmittance spectra to derive a precise wavelength calibration all along the intervals, with uncertainties lower than 2 m\AA~ (see Fig. \ref{Fig:shift_evol}). A second aspect is linked to the spectral resolution. Adaptive optics is so efficient that the image of the target may be smaller than the entrance slit of the spectrograph \citep[see][]{Dorn23}. This results in a very high resolution, which is a very positive thing, but precludes the prediction of the resolving power as a function of the slit width. Among the telluric features, many are so narrow that they can be efficiently used to adjust the actual spectral resolution. E.g., in the case of the $\beta$ CMa spectra shown in Fig. \ref{Fig: BetaCMa_1} and \ref{Fig: BetaCMa_2}, the resolving power was found to be on the order of 130,000, significantly above the achievable resolution for a target filling the slit width of 0.23". Finally, again thanks to the distribution of telluric lines over many spectral intervals, it is possible to determine how the resolving power varies along an order.

\subsection{$\snita$ transmittance}

$\snita$ is a newly added species. Its quantity is obtained from a $\snita$ climatology calculated by the REPROBUS 3D chemical-transport model \citep{Lefevre94}. 
The latest version has been described in \cite{Diouf24}. The amount of $\snita$ considered in TAPAS varies with the latitude and month.
$\snita$ produces very weak lines, always less than one percent of the continuum in normal conditions. However, its absorption features extend from 300 to 700 nm, in a widely used spectral region, especially for extrasolar planet studies. It may be a contaminant in the case of very high SNR data and when ultra-weak signals are searched for. We show in Fig. \ref{Fig:NO2}, as an example, the $\snita$ absorption for ESO-Paranal in May, here for an airmass of 2. Its very irregular series of features are characterized by a wide range of width and shape. They are hard to identify without a full model.

\subsection{HITRAN 2016 to HITRAN 2020}
The update of the HITRAN version from HITRAN 2016 \citep{Gordon2017} to HITRAN 2020 \citep{GORDON2022} brings additional improvements. It is expected that the new version includes new weak lines and corrects for small wavelength shifts in lines already present. It is beyond the scope of this article to review all differences between the two versions. Instead, we illustrate two changes in Fig. \ref{Fig:HITRAN16_20}. Based on ESO VLT/ESPRESSO spectra, \cite{Allart22} describe a few cases of discrepancies between the modeled and observed positions of $\spwat$ lines, detected thanks to the P-Cygni types of profiles obtained after division of the data by the model.  The modeled line was found to be slightly red-shifted (resp. blue-shifted) with respect to the data for the first (resp. second) case. We have compared the new and previous versions of HITRAN for the same dates and directions for the two most discrepant cases observed by the authors. The results, displayed in the figure, show that the two main lines are displaced in the way to diminish the discrepancy, i.e. blue-shifted (res. red-shifted). Such corrections may be very important, in particular, in the case of radial velocity measurements. Additionally, we note a new weak line in the red part of the spectral interval in the second plot that was not present in the HITRAN 2016 version. We warn that these changes may not be the most important ones, and that changes are not restricted to $\spwat$.

\section{Using transmittance spectra for LSF retrieval}\label{findpsf}

\begin{figure*}
\centering 
\includegraphics[width=0.99\textwidth]{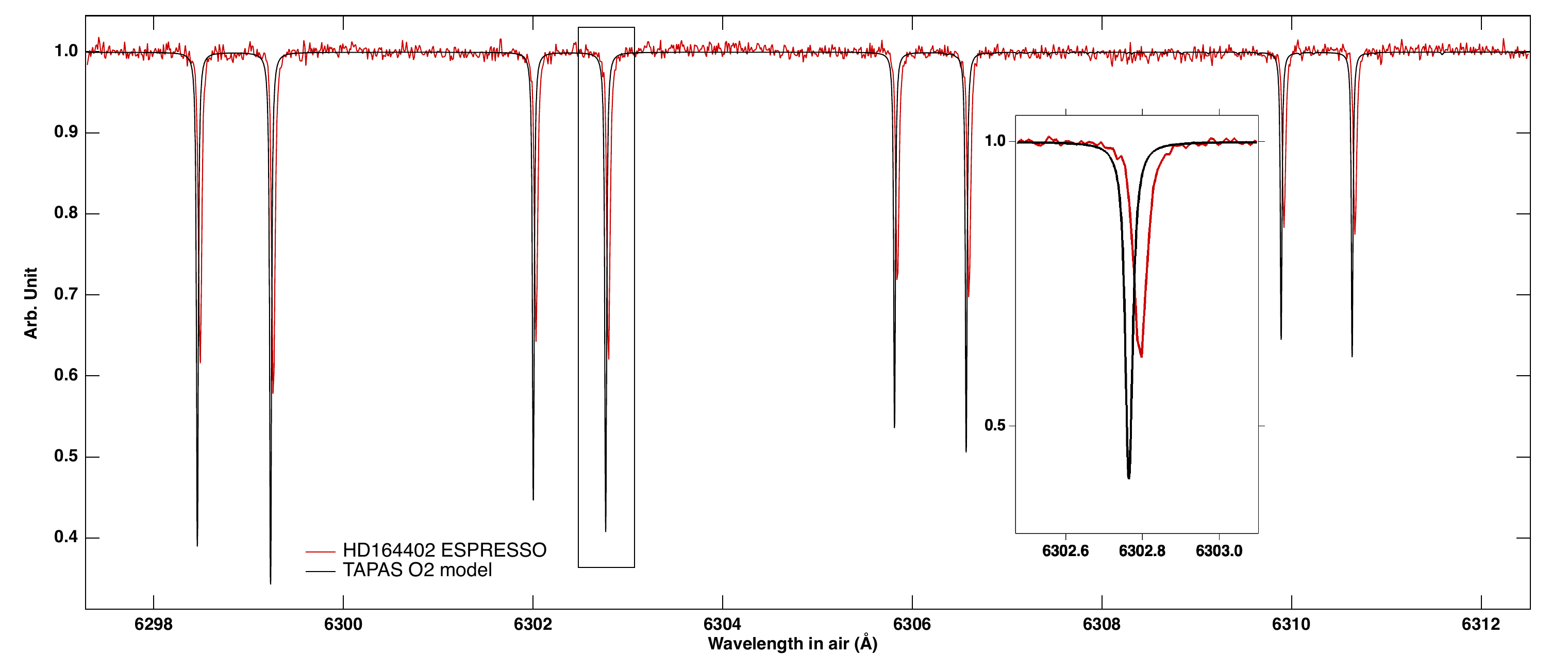}
\caption{Illustration of LSF retrieval. Superimposed are a fraction of an ESPRESSO spectrum of the hot star HD\,164402 (in red), here in the barycentric frame) and the downloaded atmospheric $\spox$ transmittance spectrum adapted to the observing conditions (in black). A zoom on one of the $\spox$ line is inserted. Note that the data are kept in the barycentric frame, as issued from the pipeline, while the TAPAS transmittance is in the geocentric frame, which explains the Doppler shift between the two. This Doppler shift will be accounted for in the computation of the LSF.}
\label{Fig:PSF_data_model}
\end{figure*}

\begin{figure}
\centering 
\includegraphics[width=0.49\textwidth]{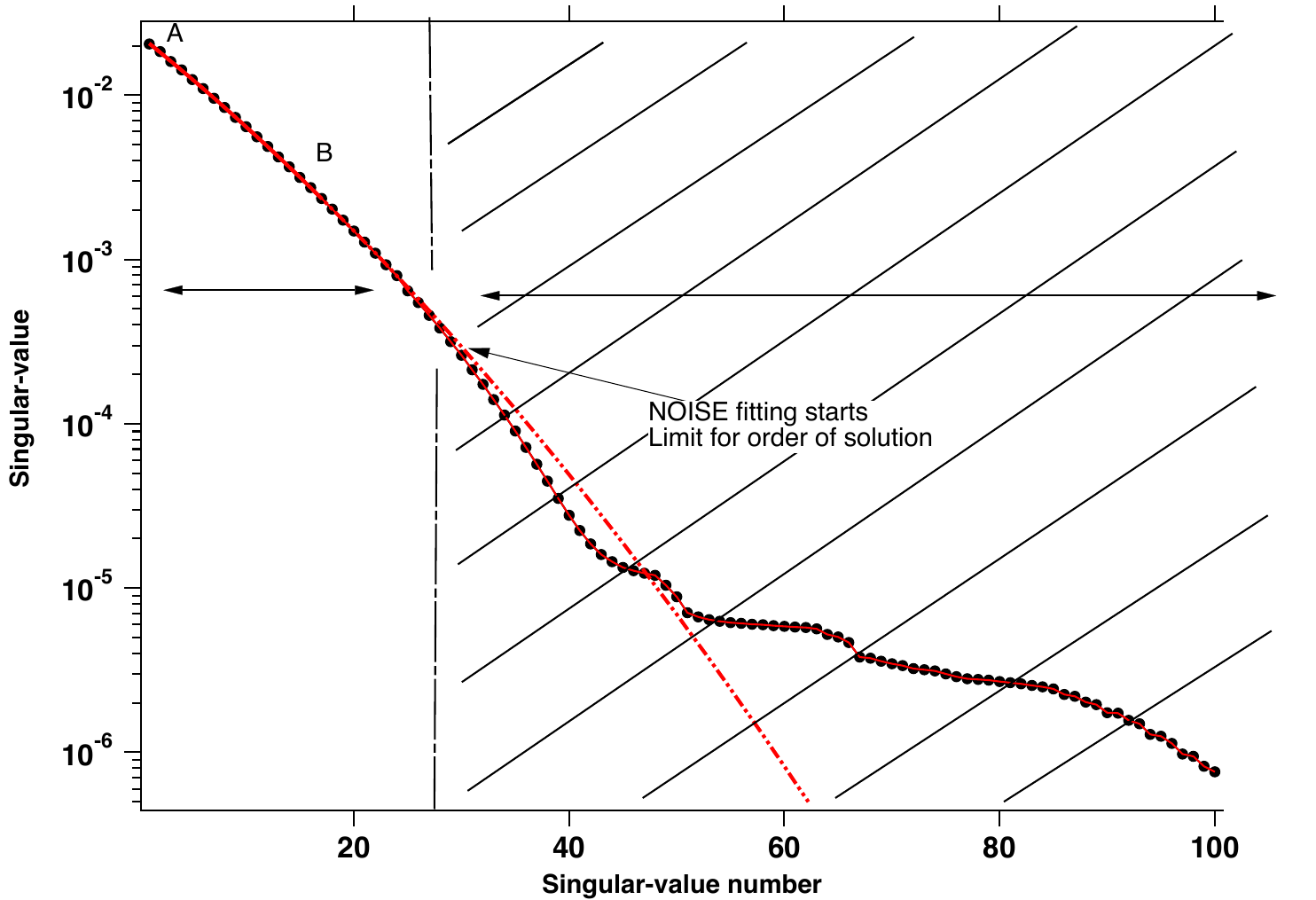}
\caption{Singular-values (SVs) of the diagonal matrix. The first value has been omitted. A second-order polynomial fit to the region delimited by A and B points is shown in red. It can be seen that the SVs start to depart from this function around a SV number of 30. Beyond this region, SVs are very low and correspond to numerical noise fitting (see text).}
\label{Fig:PSF_SVs}
\end{figure}

\begin{figure}
\centering 
\includegraphics[width=0.49\textwidth]{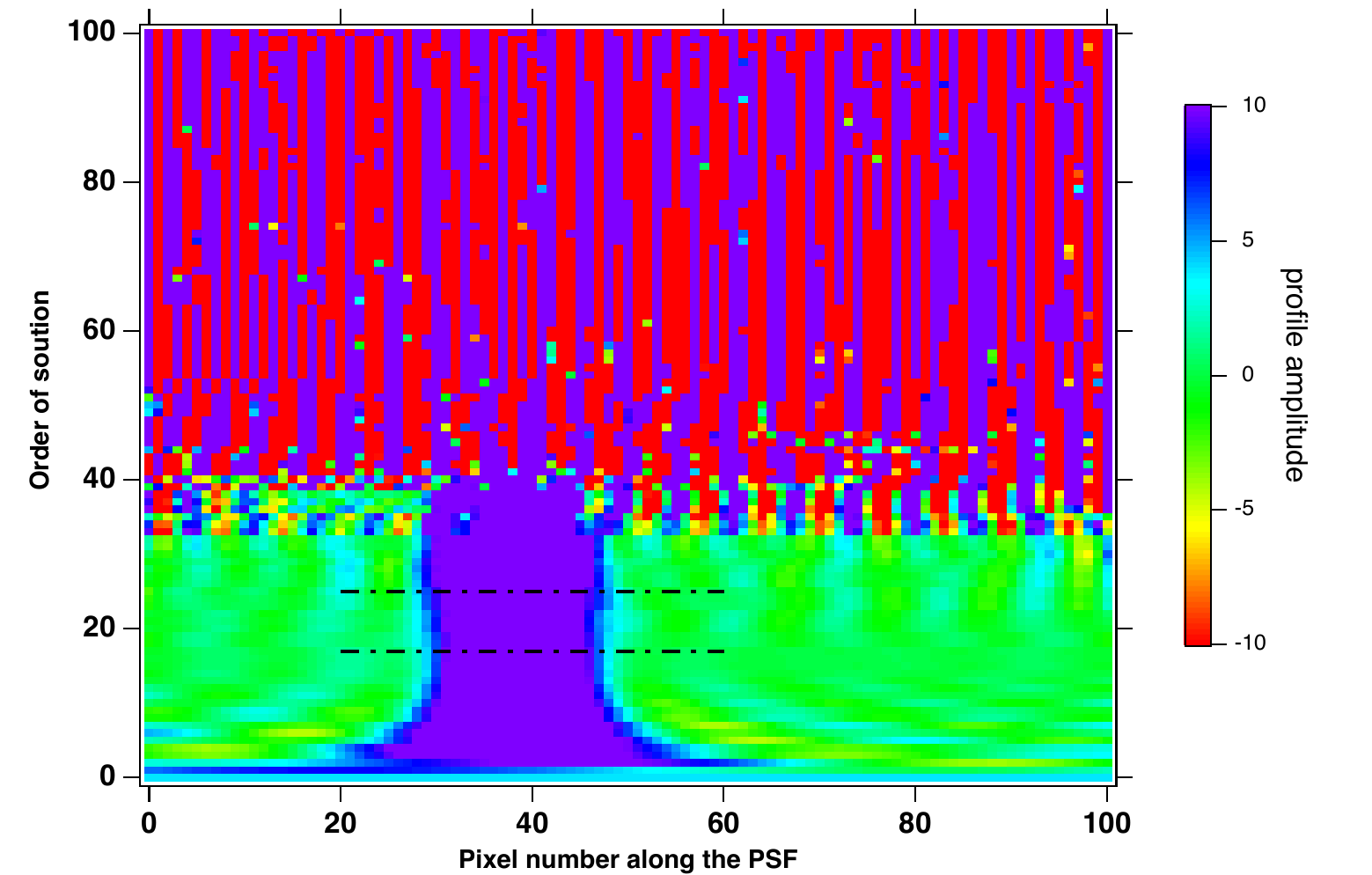}
\caption{Evolution of the LSF profile with the order of solution. Low values (at the bottom) do not reveal any acceptable shape and must be discarded. Above an order of about 30, profiles become unrealistic with appearance of negative values. They correspond to noise fitting. In-between, there is some stability around a well-defined profile.}
\label{Fig:PSF_profiles}
\end{figure}

\begin{figure}
\centering 
\includegraphics[width=0.49\textwidth]{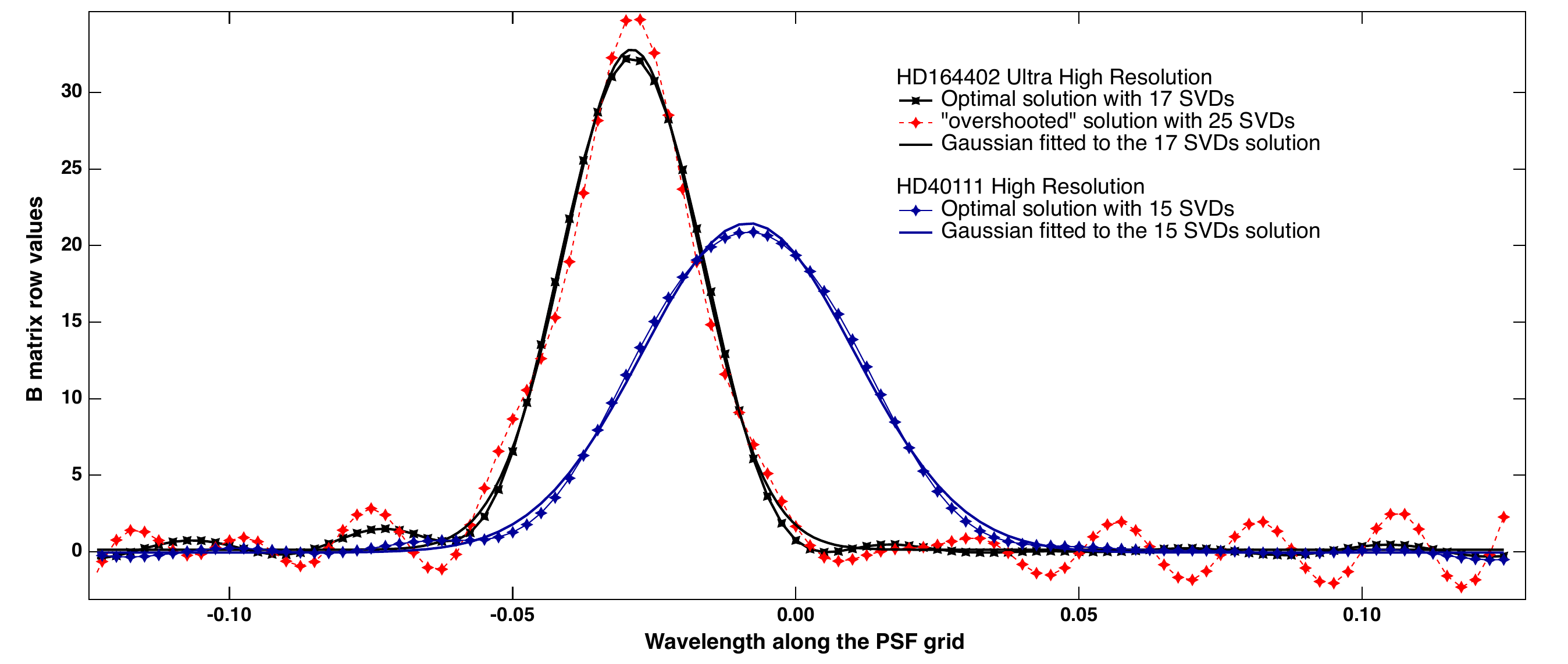}
\caption{Instrumental function profile (black curve and markers) for the ESPRESSO UHR spectrum of HD164402 and 17 SVs (corresponding to the bottom black dot-dashed line in Fig. \ref{Fig:PSF_profiles}), and the associated Gaussian fit. We show, superimposed, the profile computed for 25 SVs (red dots). An oscillatory pattern is clearly visible, suggesting that this number of SVs is already too large. Also displayed are the 15 SVs optimal profile for the ESPRESSO HR spectrum of HD40111 and its corresponding Gaussian fit.}
\label{Fig:PSF_comp_profiles}
\end{figure}

\begin{figure}
\centering 
\includegraphics[width=0.49\textwidth]{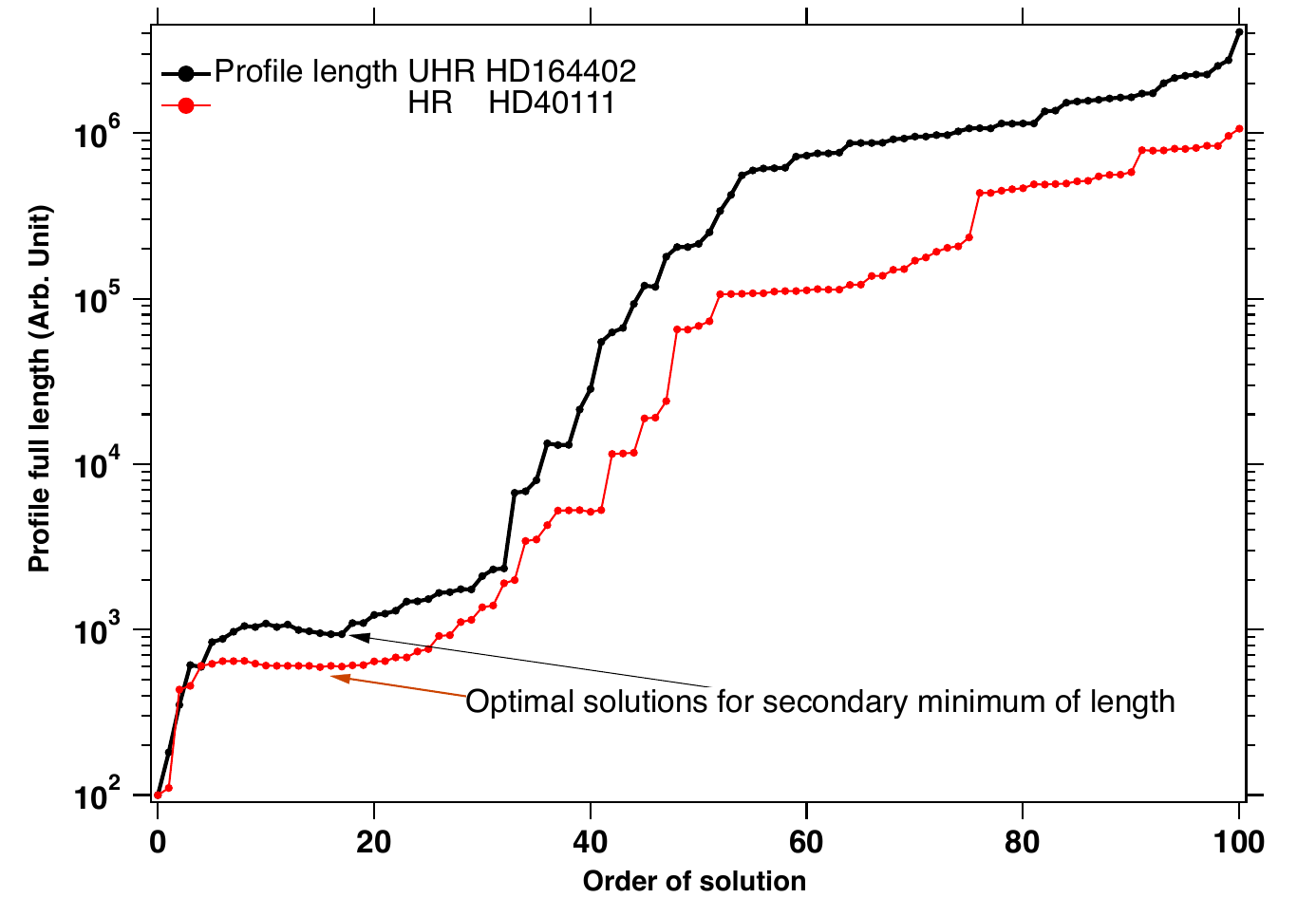}
\caption{Illustration of the criterion for the optimal number of singular values (SVs). The secondary minimum of the profile length corresponds to 17 SVs, for HD164402 (black curve), and 15 SVs for HD40111 (red curve), their respective optimal solutions.}
\label{Fig:PSF_criterion}
\end{figure}

\begin{figure}
\centering 
\includegraphics[width=0.49\textwidth]{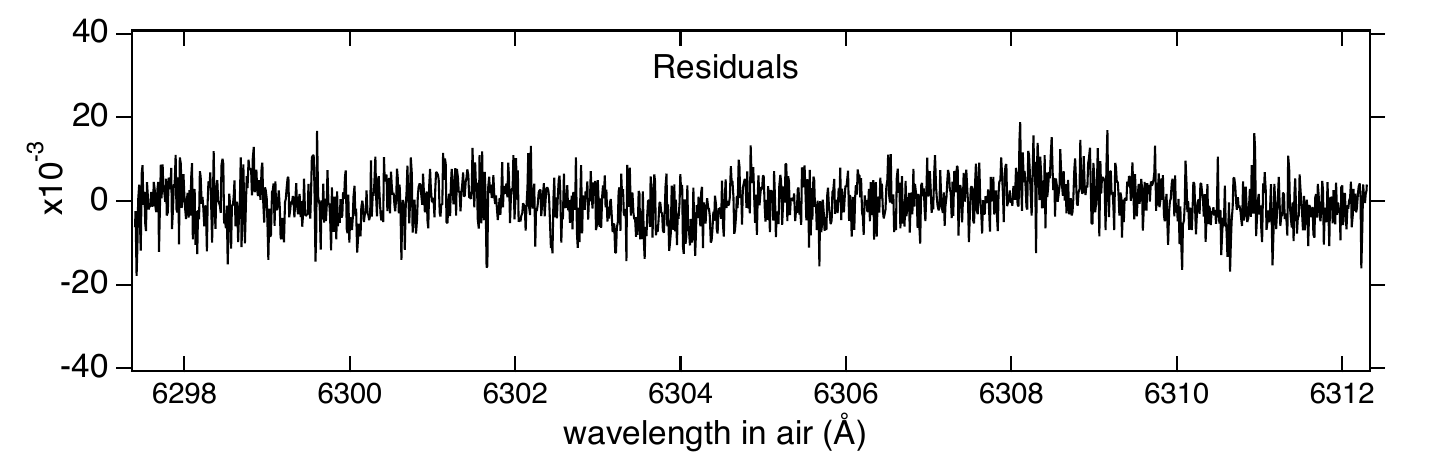}
\includegraphics[width=0.49\textwidth, height=5cm]{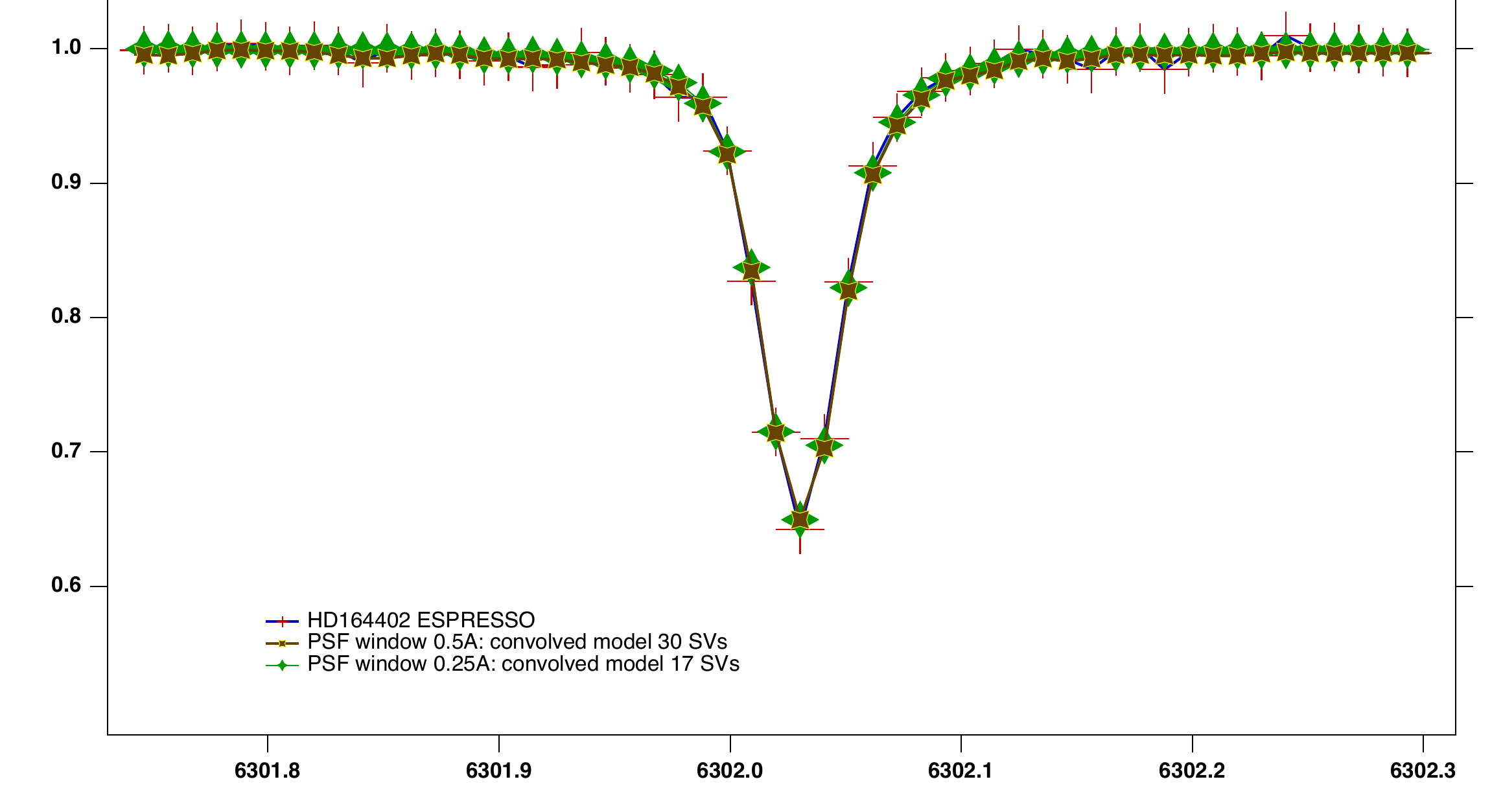}
\caption{Top: Differences between the HD164402 Ultra High Resolution data and the TAPAS model, after its convolution by the "17 SVs solution" profile. The comparison with Figure \ref{Fig:PSF_data_model} shows that there are no marked residuals at the locations of the telluric lines. Bottom: Zoom on one of the line in the UHR spectrum. In addition to the model convolved by the "17 SVs" profile (green curve and dots), the result of the convolution by the 30 SVs" profile is represented (red curve and plus signs). The two solutions are similar, however, the former solution results in a slightly more regular continuum, demonstrating its superiority.}
\label{Fig:PSF_final}
\end{figure}


Transmittance spectra provided by the TAPAS facility can be obtained with or without convolution by the LSF representing the instrumental broadening. In the latter case, the available spectral steps (e.g. $\sim$ 1.0x10$^{-3}$ \AA~ at 5000 \AA~) are much narrower than the widths of the telluric lines and all spectral features of the model are reproduced in their integrality without any broadening, or with a so-called “pseudo-infinite” resolution. 
As a consequence, they may be used in conjunction with the observed features to derive the spectral LSF of the spectrograph. In the case of an Echelle spectrograph, if $\spwat$ or $\spox$ lines are distributed over one Echelle order,  this allows to derive how the LSF evolves along the order, and refine spectral analyses.
This is particularly easy to do if the target spectrum is relatively smooth. A dedicated short exposure of a hot bright star is perfectly adapted to such an exercise. Here we illustrate the use of the telluric $\spox$ B band in spectra of two bright stars observed with the ESO ESPRESSO spectrograph \citep{Pepe21} and the adapted TAPAS $\spox$  transmittance to derive the spectrograph LSF. We make use of the method developed by \cite{Rucinski99} based on the Singular-Value Decomposition (SVD) of the matrix that contains the model and links the LSF to the data. The SVD technique was already used by \cite{Seifahrt2010} for CRIRES data. Here we describe a new method to select the optimal number of Singular Values (SVs).

\subsection{Data and Model preparation}
The LSF retrieval by means of the Rucinski's method requires the normalization of both the data and the model in the spectral region of use. In the case of the data, it can be done by masking the telluric lines, fitting the spectrum to a polynomial in unmasked regions and dividing the spectrum by the adjusted polynomial function. In the case of the model, the situation is different for $\spwat$ lines and $\spox$ bands. In the case of $\spwat$ lines the model is already normalized everywhere and there is no necessary modification. In the case of $\spox$ bands, the collisional-induced absorption (CIA) is generating a weak absorption in the continuum between the lines, resulting in a transmittance slightly below one. In this case, the same adjustment as for the data has to be performed. The second condition is that the equivalent widths (EWs) of the observed and modeled absorption features must be identical. To do so, it is sufficient to elevate the normalized transmittance spectrum at power X, X being the ratio between the equivalent width of the lines in the data and the equivalent width of the corresponding lines in the model.

We first chose as data the ESPRESSO spectrum of the B0 Ib star HD\,164402, observed in the ultra-high resolution mode R$\sim$190,000, and as spectral region a fraction of the $\spox$ B band including 8 strong lines in a region devoid of $\spwat$ lines. Figure \ref{Fig:PSF_data_model} shows data and the TAPAS model after their normalization and adjustment of EWs in this spectral interval.  We downloaded this $\spox$ model for the following parameters: 
- Observing site= ESO Paranal ; Date for the atmospheric model: 2019-03-17 at 09H33 TU; no choice of any spectral resolution 
(to get the actual unconvolved transmittance) ; $\spox$ lines solely; no Rayleigh extinction.
Since the target was observed at an airmass very close to 1, we selected a null zenith angle. It is also possible to enter the target coordinates if one does not know the zenith angle corresponding to the observations. The data are kept in the barycentric frame and the required TAPAS is geocentric. There is no need to convert data or model to adapt to a unique reference frame, the Doppler shift between the telluric features in data and their corresponding features in the model will be accounted for in the computation of the LSF.

\subsection{Algorithm}

Here we give a short description of the principle of the LSF retrieval. Details of the technique and of its advantages over other methods can be found in \cite{Rucinski99}. Mathematically, the problem is to find a solution to an over-determined linear system of n equations and m unknowns with n$>$m:
\begin{equation}
\mathbf{Y} = \mathbf{D} . \mathbf{X}
\end{equation}
where $\mathbf{X}$ is a column vector containing the LSF profile on a discretized grid with m points, $\mathbf{Y}$ is a column vector containing the observed spectrum on n points. $\mathbf{D}$ is the so-called Design Matrix, where the n lines are constituted by a series of m points of the TAPAS spectrum, shifted by some values at each successive line, according to the wavelength values of the Y vector. 
The technique proposed by \cite{Rucinski99} is a solution of these linear equations, using the Singular Value Decomposition (SVD) \citep{Press86}. It consists of decomposing the design matrix $\mathbf{D}$ into the product of 3 matrices, $\mathbf{D}= \mathbf{U} . \mathbf{W} . \mathbf{V}^{T}$, where $\mathbf{U}$ and $\mathbf{V}$ have orthonormal columns, $\mathbf{V}^{T}$ is the transposed $\mathbf{V}$, and $\mathbf{W}$ is a square diagonal matrix containing m “singular values” w$^{i}$, by decreasing order of magnitude.  It implies the computation of the inverse matrix $\mathbf{W}^{-1}$, which is also a diagonal matrix containing the inverse 1/w$^{i}$ of the singular values. The trick is to keep only a limited number k of the largest singular values of the matrix $\mathbf{W}$, to replace by 0 the largest last m-k values 1/w$^{i}$ of $\mathbf{W}^{-1}$ and then proceed to the inversion of the system, as it is described further below. The solution is given by: $\mathbf{X} = \mathbf{V} \mathbf{W}^{-1}(\mathbf{U}^{T} \mathbf{Y})$. If too many singular values are kept, the noise in the data starts to be fitted, giving physically unrealistic wiggles in the LSF. 

We have coded this algorithm, and used it to find the LSF based on ESPRESSO data and TAPAS models described above.  In addition to data and model, the method requires the choice of two parameters, the wavelength interval over which the LSF must be calculated, which must be wide enough to include at least one strong feature, and the number of points (which must be odd) used to define the LSF itself, which must be large enough in such a way that the corresponding wavelength step is significantly smaller than the expected LSF width. Here we used 101 points distributed over 0.25 \AA~.

Fig. \ref{Fig:PSF_SVs}  displays the singular-values of the diagonal matrix computed for the data and model of Fig. \ref{Fig:PSF_data_model}. The first value is close to unity and is not shown. They decrease by 5 orders of magnitude between solution orders 1 and 100 (and 7 orders of magnitude from order zero). A change in the shape is visible at about order 30, confirmed by the departure between a second order polynomial fit adjusted between 1 and 17 and extended beyond. The meaning of this change is the appearance of non-significant SV values corresponding to noise fitting. In the Rucinski's presentation, instead of a computed model such like TAPAS, the author uses a very-high resolution spectrum of the same object to find the LSF of a lower resolution spectrum. In this case, the "noise" SVs correspond to actual noise in the flux of the high-resolution spectrum used as a model, and the change of shape is spectacular. Here, the TAPAS model does not have any actual flux noise, and only extremely small numerical noise due to limitations in the computational code, therefore the evolution of the SV along the diagonal is much less spectacular than in the case of actual noise in data. 

Fig. \ref{Fig:PSF_profiles}  displays the evolution of the computed LSF profile when the solution order increases (from bottom to top). It shows in a more visible way the change when SVs above around 30 are taken into account. The profiles become totally unrealistic with an oscillatory aspect and appearance of strong negative values. The validity region between about 10 and 30 is also made particularly clear, with an almost constant profile marked by a well-defined peak. Note that the peak is not centered, which reflects the Doppler shift between data and model.  Profiles corresponding to solution order 17 and 25 (whose locations are marked) are displayed and compared in Fig. \ref{Fig:PSF_comp_profiles}. The order 17 profile corresponds to almost flat values outside the LSF peak area, while order 25 starts showing an oscillatory behavior. Note that these oscillations have an impact on the width of the LSF, which has the consequence of making the width more uncertain. At variance with order 25, for order 17 the central region is independent of any oscillation and corresponds to the actual LSF. An easy way to select the optimal solution is to compute the length of the profile and choose the order corresponding to the minimum length in the "good" area, as illustrated in Fig. \ref{Fig:PSF_criterion}. It actually corresponds to the secondary minimum, the primary minimum being reached at the first SV. Profiles with oscillations will have a longer length and will be discarded. Such a criterion is safer than selecting the minimum width of the peak area, found, e.g. by fitting a Gaussian to the LSF, due to the perturbations by the oscillations, as mentioned above, which may slightly and spuriously decrease the width. According to the length criterion, the solution order 17 is the optimal one. 

We show in Fig. \ref{Fig:PSF_final} the TAPAS transmittance after convolution by the order 17 profile for the whole spectral interval and superimpose on the data. Data and model are indistinguishable. The difference between data and model is shown at top. There are no particular features at the locations of the oxygen lines, showing the quality of the LSF adjustment. For one of the lines shown separately, the transmittance convolved by the order 25 profile is superimposed. It is not very different, with only extremely small oscillations appearing on the blue side. The convolution by a fitted Gaussian to the order 17 profile, not shown here, does not reveal any significant difference with the use of the profile as it comes out directly from the SVD method. Both can be used for the data analysis. However, the selection of the order corresponding to the second LSF minimum length (here order 17) as illustrated in Fig. \ref{Fig:PSF_criterion} has the advantage of providing an automated criterion, and is more accurate, since, as can be seen in Figure \ref{Fig:PSF_comp_profiles}, the computed LSF is slightly different from a Gaussian, in agreement with the results of \cite{Schmidt24}.

Figures \ref{Fig:PSF_comp_profiles} and \ref{Fig:PSF_criterion} also display the optimal profile and the corresponding criterion for the High resolution (HR) ESPRESSO spectrum of the early-type star HD40111. For this spectrum, the secondary minimum of the profile length corresponds to 15 SVs. The LSF is found to be much wider than for the UHR spectrum, as expected.
If the 17 SVs profile found for HD164402 is fitted with a Gaussian, the corresponding  resolution is measured to be R (UHR) $\simeq$ 225,000. This is somewhat above the value deduced from  the top part of Fig. 11 from \cite{Pepe2021} for order 130 and pixel 3300 (for wavelength $\simeq$ 6300 \AA) which is around 210,000. However, the authors warn in a note that, in the case of the UHR spectra, their determinations of R may correspond to a lower limit. A better comparison can be obtained in the case of the HR mode and the HD40111 spectrum. In this case, the resolution deduced from the 15 SVs profile is  R(HR)=143,000. According to the digitization of the middle part of Fig. 11 from \cite{Pepe2021}, whose results for order 130 are displayed in Fig. 8 from \cite{Ivanova23}, the expected resolution around the pixel 3300  is R$\simeq$148,000, quite close to our determination.

\section{Using TAPAS}\label{usingtapas}

The advantage of TAPAS is its ease of use. We detail here all the steps.  After reading of the TAPAS presentation on the home page\footnote{https://tapas.aeris-data.fr/en/home/} , first users need to select the "request form" menu, indicate an e-mail address and choose a password. An e-mail will be sent to the indicated address, and, after verification of the address, the service will be immediately available. For each request, use the "request form". 

All items from the "request form" page must be documented. For each item, a question mark provides information on the way to fill in the required parameter. We advise users to look at this detailed information for the first use of TAPAS.  First, a list of the main astronomical observatories is proposed. In case of observations from a location that is not part of the list, users may enter the observing site longitude, latitude and altitude. Users may suggest the addition of an observatory. Second, a date must be entered and the UT time of the observation. Third, a spectral unit for the transmittance must be chosen. Three choices are proposed, nanometers in vacuum, nanometers in standard air conditions, or wave number in cm$^{-1}$. 

TAPAS proposes direct line-by-line calculations, or a convolved transmittance.  The former case is useful if one wants to measure or adjust an instrumental function, as detailed in sections \ref{spectra} and \ref{findpsf} and/or if recorded spectra are at very high resolution.  The second option allows users to obtain a transmittance adapted to any chosen spectral resolving power, based on a Gaussian instrumental function.  For this second option, a sampling ratio must be indicated. It allows the user to adapt the wavelength (or frequency) step of the model to the actual pixel size of the observation, and reduce the length of the downloaded file. Then a spectral range must be entered. 

In addition to observing location and date, users are offered two choices, namely a zenithal angle or the equatorial coordinates of the target. The target coordinates are necessary if one wants to obtain a transmittance in the barycentric frame (see below). The following step is the choice of atmosphere model. For recorded data, users are advised to use Arletty, the atmosphere interpolated in the ECMWF model. In this case the TAPAS request should be sent at least two or three days after the observing date. In other cases, estimates of atmospheric transmittance in average conditions, namely six atmospheric standard models, are proposed. 

The case of water vapor is particular. Humidity conditions vary rapidly, on time scales of hours or even minutes, and ECMWF computations may not follow such rapid variations. If the observatory provides a measurement of the water vapor vertical column (usually determined by means of a radiometer) at the time of the observations, the user is advised to enter this value in the request form. In this case, TAPAS will keep the shape of the vertical profile computed by ECMWF, but scale water vapor densities to match the on-site measurement. We have tested this tool and found that it provides water vapor lines quite close to the observed ones. However, when exposures are not very short, and,  especially in the case of low water vapor pressure, because radiometer measurements do not perfectly predict atmospheric columns, the amount of water vapor deduced from the radiometer may not perfectly correspond to the actual value \citep[see, e.g. ][]{Ivanova23}. It seems that it applies, in particular, to the situations of extremely low water vapor, as we found in section \ref{spectra} in the case of CRIRES data. One way to solve this problem is to download the water vapor transmittance independently of the other species (see below), and to modify the transmittance during  a post-processing by elevating it to the power of a free coefficient $\alpha$ until $\spwat$ modeled lines coincide with the data. Given the large number of water vapor lines, the coefficient $\alpha$ is easy to determine. 

The output parameters of TAPAS products are selected in the second column of the "request form" page. Several formats are proposed, including FITS, VO, ASCII, and NetCDF. The transmittance spectra can be obtained in the observing location frame (do not select BERV from the menu in this case) or in the barycentric frame (select BERV) to allow a direct comparison with pipeline products using this frame. The various contributions to the transmittance, i.e. the Rayleigh extinction and the absorption by the proposed seven species must then be selected.  Unless one is interested in the detailed shape of the spectrum, the Rayleigh extinction can often be skipped, since it is very smooth and can be considered as part of the continuum. Finally, an important choice is the last one: the transmittance spectra can be obtained separately or in a unique file after having merged. The user is advised to select the former case, which offers the possibility of adjusting the quantity of water vapor independently of the other species. 

Frequently, users need data for series of dates or series of spectral intervals for the same observatory. TAPAS allows duplicating all the parameters of a request (use the "ADD" menu item) and modifying one or more parameters in the added request. All the requests are executed as part of a unique submission, and all results are stored in the same directory. The number of combined requests is not limited, provided the execution time for the full series of requests is compatible with the system, a condition most often achieved. 

\section{Conclusions and perspectives}

TAPAS allows to obtain online and in a very short time atmospheric transmittance spectra fully adapted to an observing site, a date and either a zenithal angle or target coordinates. We have described the improvements of the TAPAS tool, mainly a wavelength range evolution to reach 300 to 3500 nm, and the use of HITRAN 2020. 
The main advantage of TAPAS is that it is very easy to use and convenient for observers who want to identify spectral features, to predict the telluric contamination, to measure a spectral resolution or perform a wavelength assignment, or to download accurate models to be part of their local software. As an example, they can be used in forward models appropriate to cool stars whose spectral lines overlap telluric features, or if wavelength assignments and/or spectral resolution vary in a non-negligible way along the spectral interval to be analyzed. As another advantage, TAPAS is based on the ECMWF, the European Global atmospheric model, of very high reputation. 

We have illustrated the performances of TAPAS in the newly covered NIR domain by comparing its predictions with CRIRES spectra of a hot star. Except for water vapor, other absorbing species are predicted accurately. Concerning water vapor, we found that forcing its column to the value predicted by a local radiometer improves considerably the transmittance model, however, a final adjustment may be still required,  at least in the case of very low humidity. After this adjustment, the convolved product of the predicted transmittance spectra of the various intervening species fits very well the data. We showed that the predicted transmittance spectra in the highest resolution mode can be advantageously used to determine with high precision the instrumental function and the wavelength assignment \footnote{It is also intended to provide a Python code for the LSF retrieval with SVD method.  Currently developed by Anastasiia Ivanova, it should be available at GitHub \url{https://github.com/aeictf/Rucinski-deconvolution}}.

There are several planned improvements of the tool. One is the inclusion of additional species. Work  is in progress to include NO$_{3}$, whose absorption lines are very weak but contaminate the widely used optical part. It is also planned to include  C$_{2}$H$_{6}$ , HCN and HCl which produce very weak lines between 3320-3360, 2950-3100, and 3230-3500 nm respectively.  A second, planned improvement is the use of the ECMWF so-called "reanalysis-type" results in replacement of the "forecast-type" results, when the delay between the date of the observation and the date of the user's request to TAPAS is long enough to allow it. The re-analysis model for a given day is computed based on accumulated data before and after this day, while  the forecast model is based on data recorded prior this day. Although being already excellent, the forecast model is frequently surpassed in quality by the re-analysis model. Since the delay for the re-analysis calculation corresponds to about two months, and because many observers perform their final, detailed analyses at least a few months after the observations, this would be a frequent situation.  A third project is the association to the ECMWF of several photochemical model results from the Copernicus Atmosphere Monitoring Service (CAMS) Data Store whose inclusion would increase the accuracy of the predictions. E.g., an  extended use of the CAMS products, to include,  diurnal and latitudinal variations of species like $\scarb$ or $\smeth$ is foreseen.  From the practical point of view, the possibility will be given to the user to upload a list of parameters and obtain the results in a single file. About $\snita$, a dependence of its concentration on the solar zenith angle is part of the REPROBUS chemical-transport model. Such an additional variability should be included in the TAPAS computation, to increase the quality of the prediction. Finally, on another aspect, a potential link between columns of $\spwat$ predicted by the radiometer and the actual measured columns of absorbing $\spwat$ molecules will be investigated. If found, the predicted transmittance spectra will take this relationship into account, to obtain more realistic results.

\begin{acknowledgements}

We thank our referee Dr A. Smette for careful reading, very useful remarks, suggestions and corrections. This work benefited from the French state aid managed by the ANR under the "Investissements d'avenir" programme with the reference ANR-11-IDEX-0004 - 17-EURE-0006. We thank AERIS for maintaining the service TAPAS. ESO spectra are taken from programs '102.C-0699(A)' and '108.228B.001'.
We thank Paul Bristow from the ESO staff for providing information on the CRIRES spectra. This research has made use of the VizieR catalogue access tool, CDS, Strasbourg, France (DOI: 10.26093/cds/vizier). The original description of the VizieR service is published in 2000, A\&AS 143, 23.
\end{acknowledgements}

\bibliographystyle{aa}
\bibliography{tapasplusrev2}

\end{document}